\documentclass[a4paper,11pt]{article}
\pdfoutput=1

\usepackage{pub} 
\usepackage[utf8]{inputenc}
\usepackage[english]{babel}
\usepackage{ragged2e}
\usepackage{hyperref}
\usepackage{graphicx,epstopdf}
\usepackage{upgreek}

\title{\boldmath Commissioning and simulation of CHROMIE, a high-rate test beam telescope}

\author[a,1]{P. Asenov,\note{Corresponding author.}}

\affiliation[a]{Institute of Nuclear and Particle Physics (INPP), NCSR Demokritos,\\Aghia Paraskevi, Greece}

\emailAdd{patrick.asenov.asenov@cern.ch}

\abstract{The upgrade of the LHC to the High-Luminosity LHC (HL-LHC) is expected to increase the current instantaneous luminosity by a factor of 5 to 7, providing the opportunity to study rare processes and measure precisely the standard model parameters. To cope with the increase in pile-up (up to 200), particle density and radiation, CMS will build new silicon tracking devices with higher granularity (to reduce occupancy) and improved radiation hardness. During the R\&D period tests performed under beam are a powerful way to develop and examine the behavior of silicon sensors in realistic conditions. The telescopes used up to now have a slow readout (less than 10 kHz) for the needs of the CMS experiment, since the new outer-tracker modules have an effective return-to-zero time of 25 ns (corresponding to a 40 MHz frequency) and a trigger rate of 750 kHz. In order to test the CMS Tracker modules under the LHC nominal rate, a pixel telescope named CHROMIE (CMS High Rate telescOpe MachInE) is designed, built and commissioned at CERN for beam tests with prototype modules for the CMS Phase-2 Tracker upgrade. In this article the design of CHROMIE, the calibration of its modules, and its timing and synchronization aspects are presented, along with the first beam test results. In addition, the tracking algorithm developed for CHROMIE and a preliminary Geant4 simulation study of the telescope under beam are discussed.}


\collaboration[c]{on behalf of the CMS Collaboration}

\proceeding{21$^{\text{st}}$  International Workshop on Radiation Imaging Detectors\\
  7-12 July 2019 \\
  Kolympari, Chania, Crete, Greece}

\begin{document}
\maketitle
\flushbottom

\section{Introduction}
Demands for the probing of new physics have led to the planned upgrade of the LHC to the HL-LHC (High-Luminosity LHC). An increase of the total integrated luminosity in proton-proton collisions at the LHC by an order of magnitude is foreseen, from 300 fb$^{-1}$ over the period 2011-2023 to 3000 fb$^{-1}$ over the period 2026-2037. The instantaneous luminosity is expected to reach a peak of $7.5\times 10^{34}$ cm$^{-2}$ s$^{-1}$, yielding about 200 collisions per bunch crossing with a 25 ns bunch spacing~\cite{a}. This increase in luminosity will require for the silicon trackers in the major LHC experiments a considerably higher granularity (to reduce occupancy) and improved radiation hardness (to withstand the augmented total ionizing doses) compared to the existing ones. For these reasons significant improvements of the four large LHC detectors are scheduled for the next years. During this improvement campaign an important milestone of the CMS~\cite{cms} Phase-2 upgrade is the replacement of the entire Silicon Strip Tracker. Apart from the requisitions on occupancy and radiation hardness, it is crucial that the Tracker will provide information to the first level trigger (L1) in order to cut down the event selection rate to the proposed future L1 acceptance rate of 750 kHz. That is why the future CMS tracker will be built with modules consisting of two closely spaced silicon sensors read out by common front-end ASICs, which allow on-detector transverse momentum ($p_{\mathrm{T}}$) discrimination of tracks. These so-called ‘$p_{\mathrm{T}}$ modules’ will reduce data identifying the high-$p_{\mathrm{T}}$ tracks by coincidence of signals in both sensors~\cite{b}.

The ongoing second long shutdown of the LHC (LS2) concurs with the prototyping period of the new Phase-2 Tracker modules. Extensive beam tests of the silicon sensors and their readout electronics are necessary in order to examine the behavior of the sensors in realistic conditions. Thus, new detectors under development (usually referred to as Devices Under Test, DUT) can be tested for channel efficiency, cluster size and cross-talk between adjacent channels. During these beam tests a DUT is placed inside a complex system consisting of well known tracking modules called telescope. The tracking modules are highly segmented in order to reconstruct with high accuracy particle tracks and measure the tracking efficiency of the DUT. Existing telescopes commonly used by the CMS Collaboration use a Monolithic Active Pixel Sensor chip with an integration time of 115.2 $\upmu$s, which is equivalent to a 8.68 kHz readout frequency. However, the integration time of the Phase-2 Tracker modules (and other HL-LHC sensors) is 25 ns, which corresponds to a 40 MHz rate~\cite{a}, i.e. 4600 times the today’s CMS telescopes readout frequency. It is obvious that under these conditions the Phase-2 modules can't be tested at nominal rates with the old telescopes used by the CMS Collaboration. This is the main reason why new telescopes are being developed right now, like CHROMIE (CMS High Rate telescOpe MachInE) at CERN~\cite{d} and CHROMini at CYRCé, IPHC-Strasbourg~\cite{e},~\cite{f}, the telescopes with the highest rate compatible with up-to-date CMS-standard hardware and software.

This paper outlines the design of CHROMIE and its commissioning procedure. A brief description of the applied tracking and preliminary alignment method is described. Finally, a comparison is made between the data from a small beam test that took place in 2018 at CERN and the related output of a standalone Geant4 simulation program.

\section{Design, commissioning and tracking method}
As mentioned before, the detector modules for the Phase-2 Tracker need to be tested under beam to ensure good production quality before the next replacement of the Tracker detector. The highest rate that a Phase-2 Outer Tracker module might reach is expected to be 50 MHz/cm$^{2}$~\cite{b}. CHROMIE is designed to withstand particle rates up to 200 MHz/cm$^{2}$ with an expected resolution of the order of ~10-20 $\upmu$m~\cite{d}. 

CHROMIE is equipped with the same custom-made triggering, control and readout boards and DAQ (data acquisition) software as the CMS pixel detector. CHROMIE is set up in the beam line H6 of the test-beam North Area at CERN, where the Super Proton Synchrotron (SPS)~\cite{sps} supplies the different experiments with high-energy particles.

\begin{figure}[htbp]
	\centering 
	\includegraphics[width=0.5\textwidth,origin=c,angle=0]{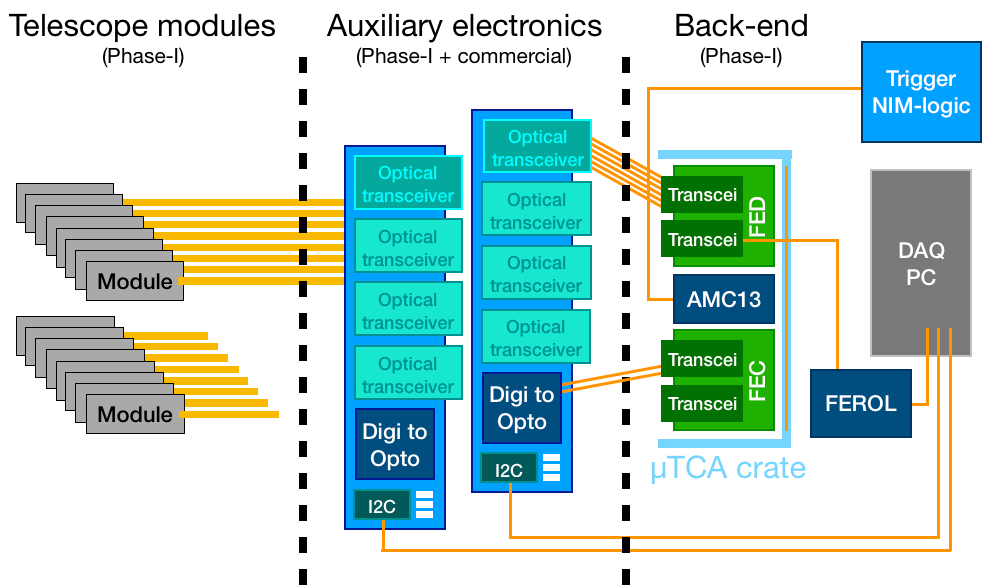}
	\qquad
	\includegraphics[width=0.4\textwidth,origin=c,angle=0]{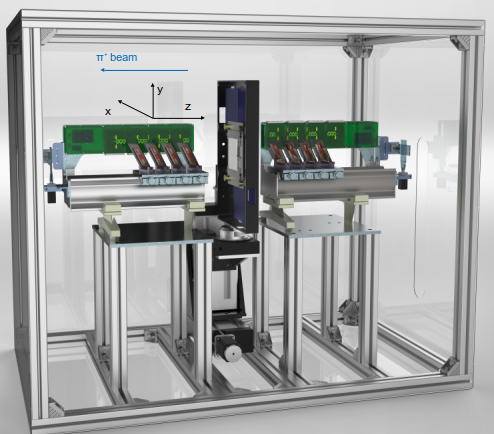}
	\caption{\label{fig:1} Left: The mostly CMS standard readout and DAQ of the CHROMIE telescope. It is analogous to the readout of the CMS Phase-1 Inner Tracker readout. The yellow ribbons represent electrical links. The orange arrows represent optical links. Right: The mechanical design of the CHROMIE telescope, with a full-size 2S module in the center as the device under test (DUT). The length of the whole box is 1.30 m. [The beam direction is right-to-left. Drawings by Nicolas Siegrist, rendering with KeyShot.]}
\end{figure}

CHROMIE consists of two "arms", each one of four layers (planes) with some dead areas, each layer containing two CMS Phase-1 BPIX (barrel pixel) modules (Grade C, with an active area of $2\times 16.2 \times 64.8$ mm$^2$) in a metal frame. One arm is in front of the DUT, and the other one behind it, on the way of the beam. The pixel size is $100\times 150$ $\upmu$m$^2$ (except for the boundaries between two read-out chips of the same module where the size in the respective direction is doubled). In order to improve the resolution and allow charge sharing between adjacent pixels, each layer is rotated by a $20^{\circ}$ tilt angle about the $x$-axis and a $30^{\circ}$ skew angle about the $y$-axis. Each layer is held by a block mounted on a carriage that can slide over rails. Auxiliary electronics is mounted close to the modules, on the rails. In between the arms of the telescope a large DUT can be placed into a box with a size of $550 \times 350 \times 40$ mm$^3$ (the cooling of which is still under investigation), with actuators for the DUT provided for translation in the $x$- and $y$-directions, while a rotation of the DUT about the $x$-axis is also allowed. Two overlapping scintillators for triggering are mounted on the rails on each end of CHROMIE and attached to photomultipliers. The trigger signal is then sent to a NIM (Nuclear Instrumentation Module) logic for shaping, discriminating and coinciding the signal with itself and a clock signal. After that, the trigger is sent to the CMS-standard AMC13 (Advanced Mezzanine Card for Slot 13)~\cite{amc} board which distributes the trigger signal from the NIM logic and the 25 ns LHC clock to the units for the control of the modules and the readout of its values: the FEC (Front End Controller)~\cite{fec} and FED (Front End Driver)~\cite{fed}, respectively. The AMC13, FEC and FED cards and their communication interface are all housed in a $\upmu$TCA crate. The FEC and FED are connected via optical fibers to the FEROL (Front-End Readout Optical Link), while the auxiliary electronics (motherboard, voltage and temperature probes) is connected via I$^2$C to the computers~\cite{d}. The readout and DAQ of CHROMIE are displayed in Figure~\ref{fig:1} (left), while the complete mechanical design of CHROMIE is shown in Figure~\ref{fig:1} (right).

Before being mounted on the telescope each BPIX module had to undergo a set of tests called pre-calibration, which included an IV test (for leakage current measurement), a pixel functionality test, a soldering (bump bond) test, a pulse height optimization test and a module response to radioactive sources test. After the installation of the modules in the telescope the calibration tests "timing calibration", "trigger latency of the calibration pulse calibration", "threshold calibration with calibration pulse" and "gain calibration of individual pixels" were performed before the commissioning in the beam, when the modules were made to run synchronously. The functionality and integrity of CHROMIE are confirmed as all telescope modules can run synchronously and there is a strong correlation between the hits in different telescope layers, which serves as an indication for particle tracks.

The readout software used by CHROMIE are the CMS-standard POS (Pixel Online Software) and XDAQ~\cite{h}. XDAQ allows the usage of distributed, hardware controlling applications called Supervisors, which can be run distributed and independently on different machines, thus controlling various components of the telescope readout, while ensuring remote data-taking and monitoring of the machine during operation. For the data unpacking and analysis, the CMS-standard offline software CMSSW (CMS Software)~\cite{i} is used. Moreover, a standalone tracking program based on CMSSW is developed by the CHROMIE Team, along with a standalone simulation program based on Geant4~\cite{j} for the prediction of residuals, cluster charge, cluster size and other magnitudes before the beam tests.

The preliminary tracking algorithm used in the data analysis is developed in Python and C++ and added as a separate module to CMSSW (CMS Software)~\cite{i}. It is called during the analysis of every test-beam run data and consists of a set of steps which are followed every time: Firstly, the clusters from noise hits are removed, secondly, alignment is applied, thirdly, the seeding is performed and finally, the pattern recognition is conducted. 

Knowing the accurate position of the telescope modules is crucial for optimizing the resolution. By reconstructing particle tracks from different data runs and minimizing the residuals over all the valid tracks, by translating the detectors parallel to the three axes and rotating them around the axes, the alignment procedure can be performed. The translations and rotations are usually performed following a series of iterations. However, at this point it should be mentioned that a detailed iterative alignment hasn’t been implemented in CMSSW yet, but this work is still in progress. Instead of the iterative alignment, only a coarse alignment is applied in our tracking algorithm, consisting of a translation of the telescope modules +50 cm parallel to the $x$-axis in global coordinates. It is demanded that all tracks should be parallel to the beam axis (the $z$-axis).

The seeding in our method is basically a search over clusters (conducted using global coordinates) for 2 points, one in the first Seeding Layer SL1 and one in the second Seeding Layer SL2 (where SL1 and SL2 run through various combinations of CHROMIE layers, e.g. the first and second Layers: L1-L2), with $\Updelta$x < 0.1 cm and $\Updelta$y < 0.1 cm. During the seeding procedure initially two $detid$ (detector ID) iterators are defined and it is checked that the first $detid$ corresponds to one of the two modules of SL1, and that the second $detid$ corresponds to one of the two modules of SL2. Loops are executed over the clusters of these modules, and then a subsequent check for clusters from noise hits follows. Afterwards, the (non-iterative) alignment is applied. Finally, the conditions for $\Updelta$x, $\Updelta$y are checked.



The seeding is initially applied on the combination of Layers L1-L2, then on L2-L3, and finally on L3-L4, until a seed is found. Further combinations are not examined, since on the layers of the arm of CHROMIE behind the DUT on the way of the beam (the second arm) two dead and one noisy modules are located.

For the pattern recognition the program looks for the cluster with the smallest distance (in two dimensions) from the track within the telescope layer, then the track is fitted including the new cluster in the list minimizing the 2D distance in the telescope layer. Short tracks which don't correspond to hits in at least 4 modules are not considered valid tracks in our algorithm.

After performing the tracking algorithm on a run of 32536 events (run 100368) from a beam test with 120 GeV $\uppi^{+}$, which has taken place at CERN, it is found that:

\begin{equation}
\label{eq:1}
\begin{split}
\frac{Number\: of\: events\: with\: at\: least\: 4\: layers\: with\: at\: least\: 1\: cluster\: and\: 0\:seeds}{Number\: of\: events\: with\: at\: least\: 4\: layers\: with\: at\: least\: 1\: cluster}
\\= \frac{6082}{28551} = 21.3 \%  \
\end{split}
\end{equation}

From \eqref{eq:1} it is found that the seeding efficiency is $78.7 \%$. Furthermore, since the number of events with 0 layers with at least 1 cluster  is 1015 in that run and $1015/32536 = 3.12 \%$, the performed analysis on run 100368 gives us a good estimation at $96.88 \%$ for the upper limit of the efficiency.

\section{Test-beam data and simulation comparison}
Subsequently, a comparison between the test-beam data and the Geant4 simulation results is presented. All the plots from the beam test are derived from the analysis of the aforementioned run 100368 with 32536 events, where the beam diameter is estimated from the simulation to be 15 mm, while only the left modules of each telescope layer (as seen on the way of the beam) are hit. As there are 2 inactive modules on the left side of CHROMIE (in L6 and L7) only 6 out of 8 modules are used in the experimental data analysis, contrary to the 8 out of 8 modules used in the simulation. Of those six modules one is significantly noisier than the rest in the analyzed run.

Different ionizations, Bremsstrahlung, pair production, annihilation, the photoelectric effect, gamma production, Compton scattering, Rayleigh scattering and the Klein-Nishina model for differential cross section calculations are among the processes included in the physics list~\cite{j} of the standalone Geant4 program for CHROMIE. The chosen statistical parameters of the general particle source (GPS) for the 120 GeV $\uppi^{+}$ are: $\sigma = $ 100 keV; position $= (-0.3,\; -0.65,\; 200)$ cm; the shape of the beam is selected ellipsoidal with $x_{half} = y_{half} = $ 7.5 mm. For each event the program calculates the stored energy in each pixel on each of the telescope modules and when dividing this energy by the one required for a single electron-hole pair production in silicon ($=$ 3.67 eV) one can find the charge (in number of electrons) collected in each pixel. If this charge exceeds the threshold of the Phase-1 pixels (set at 1700 electrons) it is considered that there is a hit in the examined pixel in the current event. Charge sharing between adjacent pixels has also been included in the simulation, taking into account weight factors resulting from the pulse heights corresponding to the different collected charge in each pixel. The impact point of the beam in each module is defined as the mean point of the centroids of the front surfaces of the hit pixels multiplied by their corresponding weight factors. A line of best fit is then calculated for the impact points in all the layers, which is afterwards used in the estimation of the residuals.

\begin{figure}[htbp]
	\centering 
	\includegraphics[width=1.0\textwidth,origin=c,angle=0]{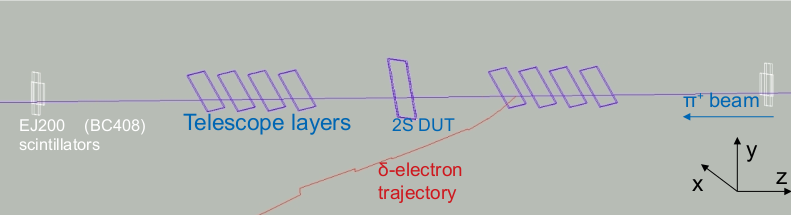}
	\caption{\label{fig:3} Visualization of the Geant4-simulated geometry of CHROMIE under beam. The DUT is a 2S module: 2 Si sensors (102700 $\upmu$m $\times$ 94108 $\upmu$m $\times$ 320 $\upmu$m), with spacing between the sensors: 2 mm; strip pitch: 90 $\upmu$m; active depth: 240 $\upmu$m.}
\end{figure}

The visualized geometry of the Geant4 simulation with a primary $\uppi^{+}$ track, a secondary $\updelta$-electron track and a 2S module used as a DUT is shown in Figure~\ref{fig:3}. From the simulation, the angular straggling is estimated at about 50-60 $\upmu$rad on average.

\begin{figure}[htbp]
	\centering 
	\includegraphics[width=0.6\textwidth,origin=c,angle=0]{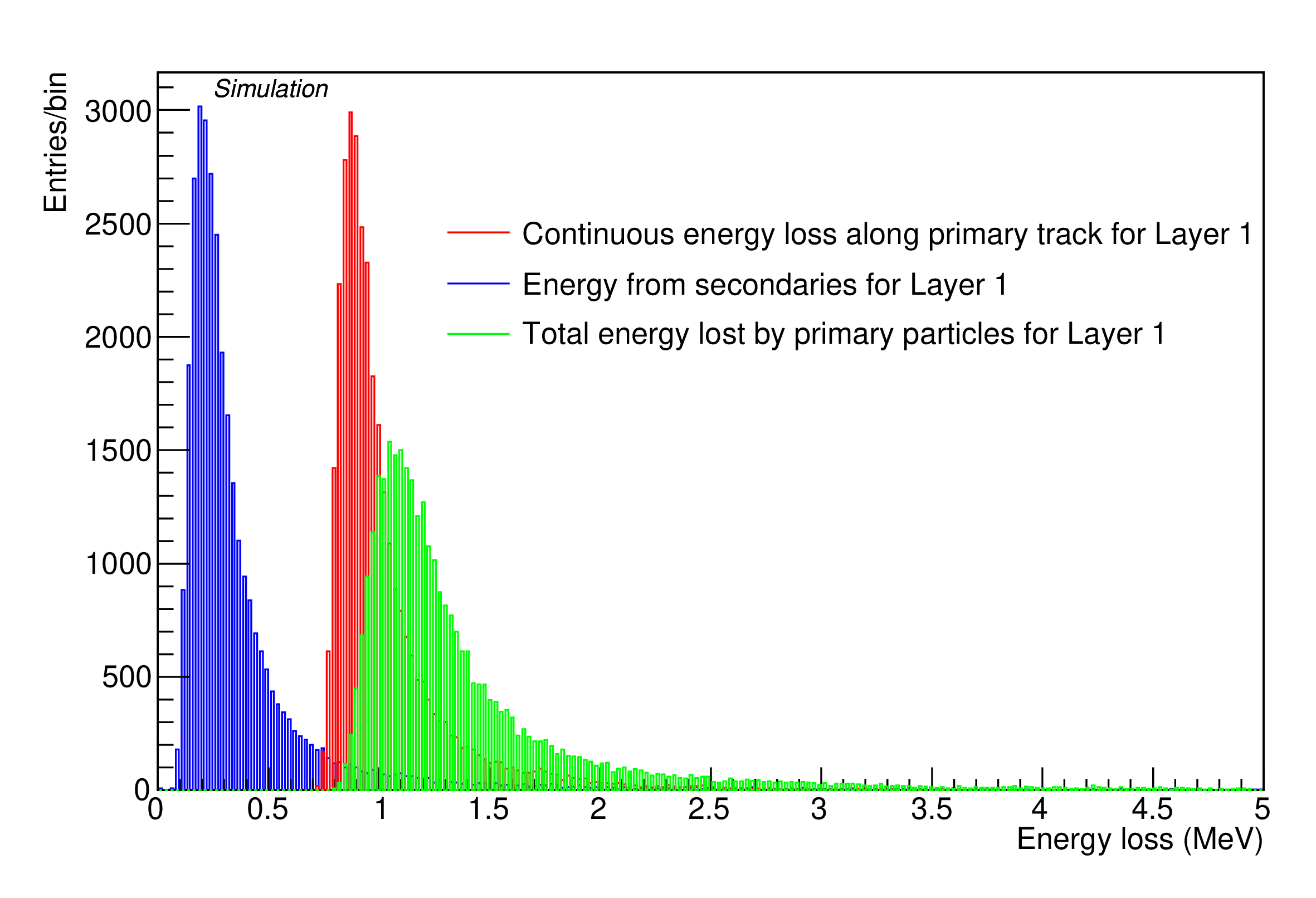}
	\caption{\label{fig:4} Energy lost by primary particles (120 GeV $\uppi^{+}$) in CHROMIE Layer 1 (simulation).}
\end{figure}

The total energy lost by the primary pions of the beam inside the silicon volume of CHROMIE Layer 1 is given in Figure~\ref{fig:4} as the sum of the energy deposited in the material and the kinetic energy of newly produced secondary particles (mostly $\updelta $-electrons), with a most probable value (MPV) equal to 1.086 MeV and $\upsigma = $ 0.098 MeV after a Landau fit.


\begin{figure}[htbp]
	\centering 
	\includegraphics[width=0.6\textwidth,origin=c,angle=0]{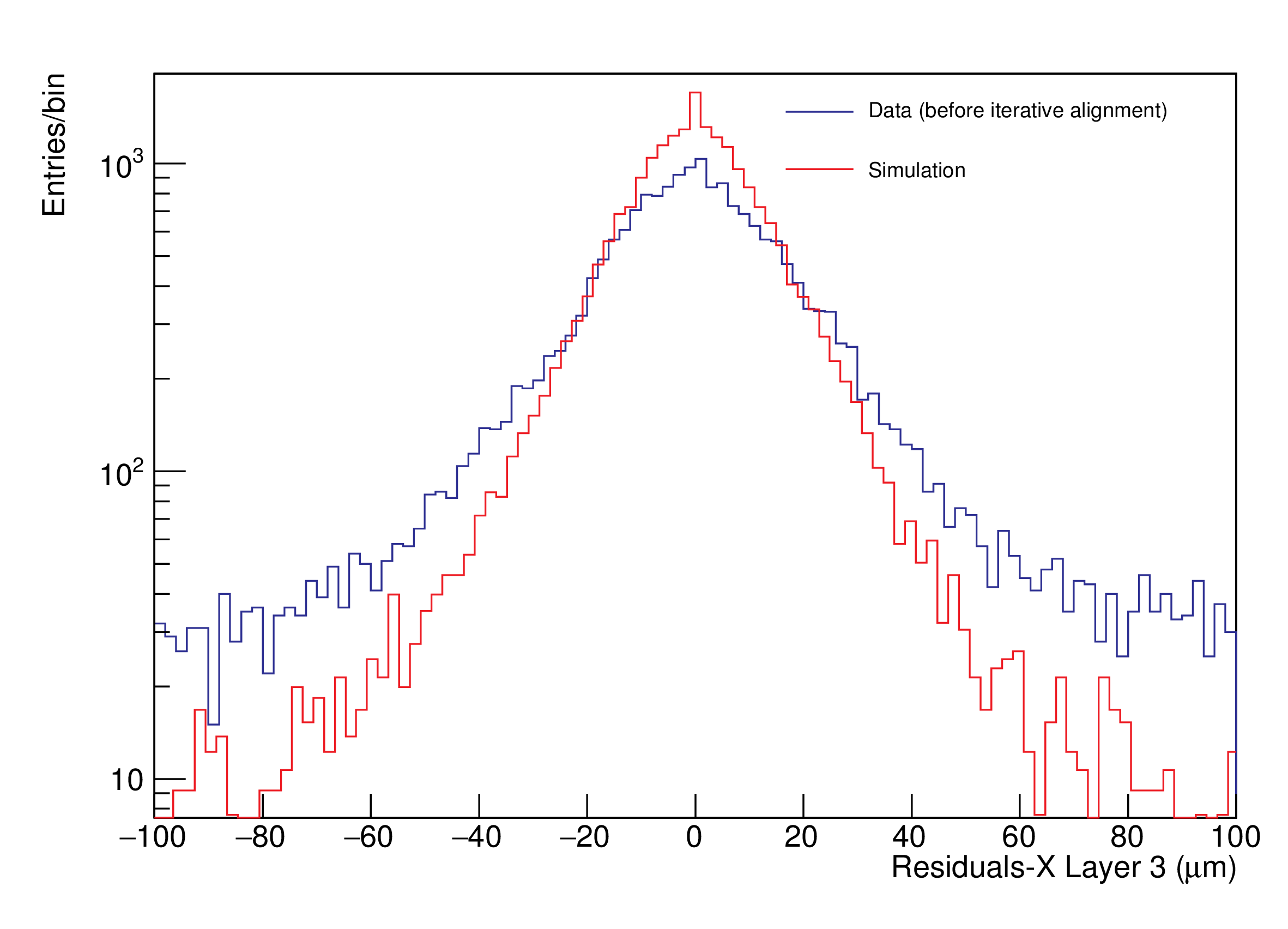}
	\caption{\label{fig:6} X-residuals for the left module of Layer 3 for a 120 GeV $\uppi^{+}$ beam (comparison between beam test data before iterative alignment and simulation). Simulation residuals scaled for 26814 valid tracks from the beam test run out of 32536 total events. \newline Standard deviation (beam test): 29.1 $\upmu$m; standard deviation (simulation): 19.8 $\upmu$m.}
\end{figure}

As seen in Figures~\ref{fig:6} the residuals in the $x$-direction of Layer 3, respectively, are larger in the test-beam data than the predicted ones. (This is also valid for the residuals in the $y$- direction of Layer 3, where their standard deviation in the beam test data is 31.4 $\upmu$m, while the standard deviation in the simulation is 23.3 $\upmu$m.) The same behavior is observed in the rest of the modules where sometimes the standard deviation might even exceed 30 $\upmu$m. This is mostly due to the oversimplified one-step alignment method. The residuals are expected to improve after the implementation of iterative alignment.

\begin{figure}[htbp]
	\centering 
	\includegraphics[width=0.47\textwidth,origin=c,angle=0]{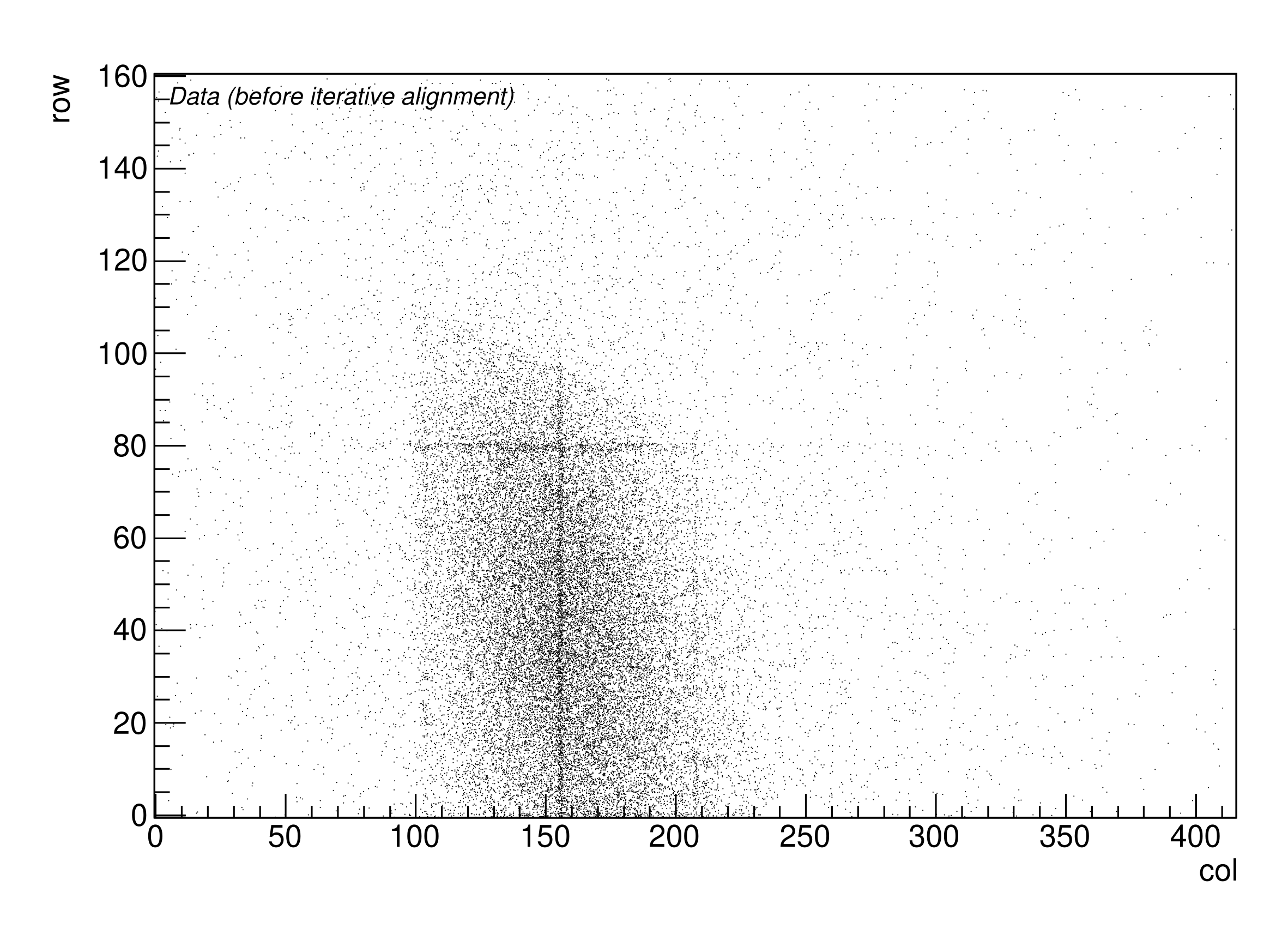}
	\qquad
	\includegraphics[width=0.47\textwidth,origin=c,angle=0]{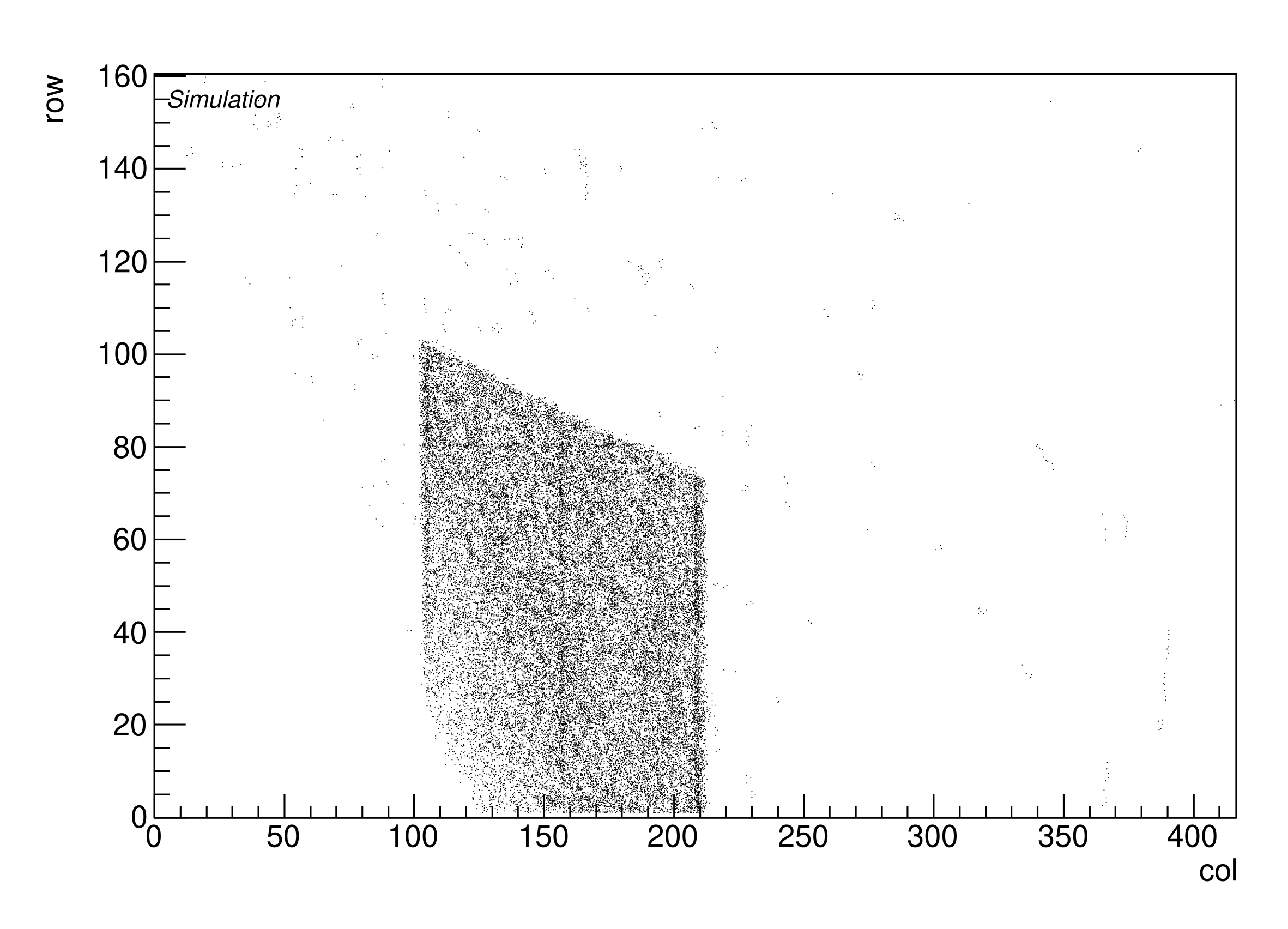}
	\caption{\label{fig:8} Hit positions per column per row for the left module of Layer 2 for a 120 GeV $\uppi^{+}$ beam (left: beam test, right: simulation with beam diameter = 15 mm and $\sigma_{E} =$ 100 keV); the beam size is measured from the beam spot on the hit position map for the same module, obtained from the analysis of the real run, and thus the above parameters are selected for the simulation run.}
\end{figure}

Figure~\ref{fig:8} depicts the hit positions for the left module of Layer 2. By counting the size of the beam spot on the module in pixels (from the test-beam plot) and multiplying it by the pixel size in each direction, the value of the beam diameter is estimated, and thus it is introduced as parameter in the simulation run. The shape of the beam spot can be explained by the rotations of the modules initially about the $x$-axis and subsequently about the $y$-axis. The noisy pixels are clearly visible. The related test-beam and simulation plots are similar for all the other layers of CHROMIE.

In addition, a cluster size investigation is made for all the modules in the $x$- and $y$- directions. The mean cluster size in $x$ in number of pixels derived from the beam test data lies in the range [1.469, 1.991] for the different modules, where the average value for all modules is 1.875, compared to the range [1.639, 1.669] extracted from the simulation, where the average value for all modules is 1.654. On the other hand, the mean cluster size in $y$ in number of pixels derived from the beam test data lies in the range [1.355, 1.754] for the different modules, where the average value for all modules is 1.665, compared to the range [1.577, 1.608] extracted from the simulation, where the average value for all modules is 1.596. The slight differences are due to the presence of broken and noisy pixels.

\section{Conclusions}
A new high-rate telescope has been commissioned at CERN, mostly based on technology developed for the CMS experiment, thus being compatible with previously existing CMS hardware and software. It can be used for tests of front-end (FE) electronics under high particle rate and high occupancy, to study the performance and saturation effects vs. track rate, and to monitor effects of radiation damage e.g. on silicon sensors. A standalone simulation program based on the Geant4 toolkit has been developed to predict the response of the telescope under various types of particle beams, which could be used as a potential base for future simulations of any particle telescope. This program could be used for estimating unknown beam parameters through comparison of its output with plots from real data where some magnitudes are unknown. There is currently a good comparison in the resolution and cluster occupancy between the beam test data and the simulation results, and further improvement is expected after the development of an iterative alignment method.

\acknowledgments
The author would like to acknowledge the support by the Hellenic Foundation for  Research and Innovation, HFRI (Greece).

The author also acknowledges the contributions of his colleagues from the CHROMIE team: Bora Akg\"un,  J\'er\'emy Andrea, Caroline Collard, Nikkie Deelen, Sandro Di Mattia, Gabrielle Hugo, Tivadar Kiss, Aristoteles Kyriakis, Dimitrios Loukas, Stefano Mersi, Nicolas Siegrist, Tam\'as T\"olyhi, Andromachi Tsirou, Viktor Veszpr\'emi. In addition, the author would like to thank Imtiaz Ahmed, Eric Albert, Jonathan Fulcher, Dominik Gigi, Jean-Fran\c{c}ois Pernot, Hans Postema and Piero Giorgio Verdini for their support!



\begin{thebibliography}{99}
	
\justifying
		
\bibitem{a}
G. Apollinari et al., \emph{High-Luminosity Large Hadron Collider (HL-LHC): Technical Design Report V. 0.1}, \emph{CERN Yellow Rep.Monogr.} {\bf 4} (2017) 1-516 CERN-2017-007-M.

\bibitem{cms}
CMS Collaboration, \emph{JINST 3 S08004 (2008)}.

\bibitem{b}
CMS Collaboration, \emph{The phase-2 upgrade of the CMS tracker}, \emph{CERN-LHCC-2017-009, CERN, Geneva Switzerland} {\bf [CMS-TDR-17-001]}.

\bibitem{d}
N. Deelen, S. Mersi, \emph{CHROMIE - The CMS High Rate Telescope}, \emph{7th Beam Telescopes and Test Beams Workshop} {\bf 14-18 January 2019}, CERN, \href{https://indico.cern.ch/event/731649/contributions/3237194/} {https://indico.cern.ch/event/731649/contributions/3237194/}.

\bibitem{e}
P. Marchand, A. Ouadi, M. Pellicioli, D. Brasse, \emph{Cyrcé, un cyclotron pour la recherche et l’enseignement en Alsace}, \emph{L’Actualité Chimique} {\bf 386} (2014) 9-14.

\bibitem{f}
P. Asenov, \emph{Test beam facility at CYRCé for high particle rate studies with a CMS upgrade module: design and simulation}, \emph{7th Beam Telescopes and Test Beams Workshop} {\bf 14-18 January 2019}, CERN, \href{https://indico.cern.ch/event/731649/contributions/3237181/} {https://indico.cern.ch/event/731649/contributions/3237181/}.

\bibitem{sps}
J.B. Adams, \emph{A design of the European 300 GeV research facilities}, \emph{MC-70-60; CERN-SPC-299, CERN, Geneva Switzerland, 1970}.

\bibitem{amc}
E. Hazen, A. Heister, C. Hill, J. Rohlf, S.X. Wu, D. Zou, \emph{The AMC13XG: a new generation clock/timing/DAQ module for CMS MicroTCA}, \emph{JINST 8 (2013) C12036}.

\bibitem{fec}
K. Kloukinas, W. Bialas, C. Ljuslin, A. Marchioro, E. Murer, C. Paillard, F. Drouhin, E. Vlasov, \emph{FEC-CCS: A common front-end controller card for the CMS detector electronics}, \emph{C06-09-25.12}, Sep 2006, p.179-184.

\bibitem{fed}
S. A. Baird et al., \emph{The CMS Tracker Readout Front End Driver}, \emph{8th Workshop on Electronics for LHC Experiments}, Colmar, France, 9 - 13 Sep 2002, pp.296-300 (CERN-2002-003).

\bibitem{h}
\emph{XDAQ CMS Online Software project page}, \href{https://twiki.cern.ch/twiki/bin/view/CMSPublic/CMSOS} {https://twiki.cern.ch/twiki/bin/view/CMSPublic/CMSOS}.

\bibitem{i}
\emph{CMS Offline Software}, \href{https://cms-sw.github.io/} {https://cms-sw.github.io/}.

\bibitem{j}
S. Agostinelli et al., \emph{Geant4: A simulation toolkit}, \emph{Nucl. Instrum. Meth. A} {\bf vol. 506} (2003) pp. 250-303.

\end{thebibliography}
\end{document}